\documentclass[12pt,twoside,a4paper]{article} 
\usepackage{graphicx}
\begin{document}
\title{Multiple Parton Interactions Studies at CMS}
\author{Paolo Bartalini\footnote{National Taiwan University} 
and Livio Fan\`o\footnote{INFN and Universit\`a degli Studi di Perugia}\\
{\it on behalf of the CMS collaboration}}
 
\maketitle

\begin{abstract}
This paper summarizes the Multiple Parton Interactions studies in CMS, focusing on the already performed low $p_T$ QCD measurements up to highest centre of mass energies of $7$ TeV and discussing the plans for the direct measurement of the multiple high-$p_T$ scatterings.

The underlying event in pp interactions is studied measuring the charged multiplicity density and the charged energy density in the transverse region, which is defined considering the azimuthal distance of the reconstructed tracks  with respect to the leading track-jet of the event, defined from tracks according to a jet clustering algorithms.

In addition, we present the measurement of the underlying event using the jet-area/median approach, demonstrating its sensitivity to different underlying event scenarios. 

Observations in the central region are complemented by the measurement of the energy flow in the forward direction for minimum bias and central di-jet events. 

We compare our underlying event and forward results with the predictions from different Monte Carlo event generators and tunes, aiming to best parametrize the multiple parton interaction energy dependence starting from the Monte Carlo tunes developed to best fit the charged particle spectra measured at central rapidities. 

Finally we discuss the strategy to directly measure the multiple particle interactions rate focusing on the topologies with two hard scatterings in the same event.

\end{abstract}

%\newpage

\section{Introduction}

The Multiple Parton Interactions (MPI) are currently invoked to account for observations at hadron colliders that would not be explained otherwise: the activity of the Underlying Event (UE), the cross sections of multi-jet production, the survival probability of large rapidity gaps in hard diffraction, etc. \cite{Bartalini:2010su}.
At the same time, the implementation of the MPI effects in the Monte Carlo models is quickly proceeding through an increasing level of sophistication and complexity that achieves deep general implications for the LHC physics.

%In the past, Double Scattering (DS) phenomena in the hard scattering phenomenology suggests the extension of the same perturbative picture to the soft  regime, giving rise to the first implementation of the  Multiple Parton Interaction (MPI) processes in a QCD Monte Carlo model by T.Sj$\:o$strand and M.van Zijl. Such model turned out to be very successful in reproducing the observed charged multiplicity distributions and in accounting for the violation of the sensitive Koba Nielsen Olesen scaling violation at increasing center of mass energies.

%The implementation of the MPI in the QCD Monte Carlo models is quickly proceeding through an increasing level of sophistication providing a modeling of the diffractive interactions in the same context. 

%Considerable progress in the phenomenological study of the Underlying Event (UE) in jet events is achieved  by the CDF experiment at the Tevatron collider, with a variety of redundant measurements relying both on charged tracks and calorimetric clusters, the former being intrinsically free from the pile-up effects and achieving a better sensitivity at low p$_{T}$.
%Challenging tests to the universality features of the models are provided by the extension of the UE measurement to the Drell Yan topologies and by the additional complementary measurements on MB events dealing with the correlations between charged multiplicity and average charged momentum.

This contribution summarizes the early Underlying Event (UE) and forward measurements of the CMS collaboration using pp collision data up to highest energies of $\sqrt{s}=7$ TeV. It also reports along the feasibility study for the direct measurement of double parton scattering phenomena focusing on the $3jet + \gamma$ channel. 

A detailed description of the CMS detector is available in Ref.~\cite{:2008zzk}. Generator level Monte Carlo (MC) predictions are compared to the data corrected with a bayesian unfolding technique taking into account the detector effects~\cite{agostini}. 

The predictions for inelastic events are provided here by several tunes of the  PYTHIA program, versions 6.420~\cite{Sjostrand:1986ep,Sjostrand:2006za} and 8.135\footnote{PYTHIA version 8.108 is used in the feasibility studies reported in section {\ref{feas}}.}~\cite{Sjostrand:2008vc,Corke:2009pm}. PHOJET \cite{Bopp:1998rc} is also used in the forward measurements:

The pre-LHC tune D6T~\cite{Field:2008zz,rdf1} of PYTHIA-6, which describes the
lower energy UA5 and Tevatron data, is a widely used reference that will
also be used for most of the presented analyses.
The tunes DW~\cite{rdf1} and CW~\cite{Khachatryan:2010pv}, which were found to
describe best the UE CMS data at $0.9$ TeV whereas D6T predictions were
too low~\cite{Khachatryan:2010pv}, will also be discussed for the $7$ TeV data.
The pre-LHC tune Perugia-0 \cite{Skands:2010ak} and  the new PYTHIA-6 tune, Z1~\cite{rick:Z1}, includes $p_T$ ordering of parton showers and the new PYTHIA MPI model~\cite{Skands:2007zg}. 
It implements the results of the Professor tunes~\cite{Buckley:2009bj} considering the fragmentation and the color reconnection parameters of the AMBT1 tune \cite{:2010ir}, while the CMS UE results~\cite{Khachatryan:2010pv, QCD-10-010} discussed in this paper have been used to tune the parameters governing the value and the $\sqrt{s}$ dependence of the transverse momentum cut-off that in PYTHIA regularizes the divergence of  the leading order scattering amplitude as the final state parton transverse momentum $\hat{p}_T$ approaches $0$.
The tune Z2 is similar to Z1, except for the transverse momentum cut-off  at the nominal energy of  $\sqrt{s_0} = 1.8$ TeV which is decreased of $0.1$ GeV/c.
PYTHIA-8 also uses the new PYTHIA MPI model, which is interleaved with parton showering. 
The default Tune 1 (indicated as PYTHIA-8) in and the new  PYTHIA-8 tune 4C~\cite{Corke:2010yf}, which focuses on the description of the early LHC data, are adopted here. It includes soft and hard diffraction~\cite{diffr-in-pythia_navin}, 
whereas only soft diffraction is included in PYTHIA-6; the diffractive 
contributions are, however, heavily suppressed by the trigger and event 
selection requirements, especially for large $p_T$ values of the leading 
track-jet. 
The PYTHIA-8 tune 4C also focuses on the description of the early LHC data.
The parton distribution functions used to describe the interacting protons are
the CTEQ6LL set for D6T, Z2 and 4C. The CTEQ5L set is adopted for the other 
simulations~\cite{Lai:1999wy,Pumplin:2002vw}.

%%%%%%%%%%%%%%%%%%%%%%%%%%%%%
% UE IN CENTRAL REGION
%%%%%%%%%%%%%%%%%%%%%%%%%%%%%
\section{The Early Underlying Event Measurements}

In the presence of a hard process, characterized by particles or clusters of particles with a large transverse  momentum $p_T$ with respect to the beam direction, the final state of hadron-hadron interactions can be described  as the superposition of several contributions:
products of the partonic hard scattering with the highest $p_T$, including initial and final state radiation;
hadrons produced in additional MPI; ``beam-beam remnants" (BBR) resulting from the hadronization of the partonic constituents that did not participate in other scatterings.
Products of MPI and BBR form the UE, which cannot be separated from initial and final state radiation.

The early CMS UE measurements focus on the understanding of the UE dynamics studying charged particle production with two different approaches.
The first (traditional) approach \cite{Khachatryan:2010pv, QCD-10-010} concentrates on the study of the transverse region, which is defined considering the azimuthal distance of the reconstructed tracks  with respect to the leading track or leading track-jet of the event: 
$60^\circ <  |\Delta\phi| < 120^\circ$. The jet reconstruction algorithm used in these studies is SisCone~\cite{Salam:2007xv}. 
On top of the traditional approach, a new methodology using anti-$K_T$ jets~\cite{Cacciari:2008gp}  and relying on the measurement of their area~\cite{Cacciari:2009dp} is adopted for the first time by CMS using charged particles in pp collision data collected at $\sqrt{s} = 0.9$ TeV \cite{QCD-10-005}. The new set of UE observables consider the whole pseudorapidity-azimuth plane instead of the transverse region and inherently take into account the leading jets of an event. 
% QCD-10-010 (and QCD-10-001).
%%%%%%%%%%%%%%%%%%%%%%%%%%%%%%%%%%%%%%%%%%%%%%%
\begin{figure}[htbp]
\begin{center}
\includegraphics[angle=90, width=0.4 \textwidth]{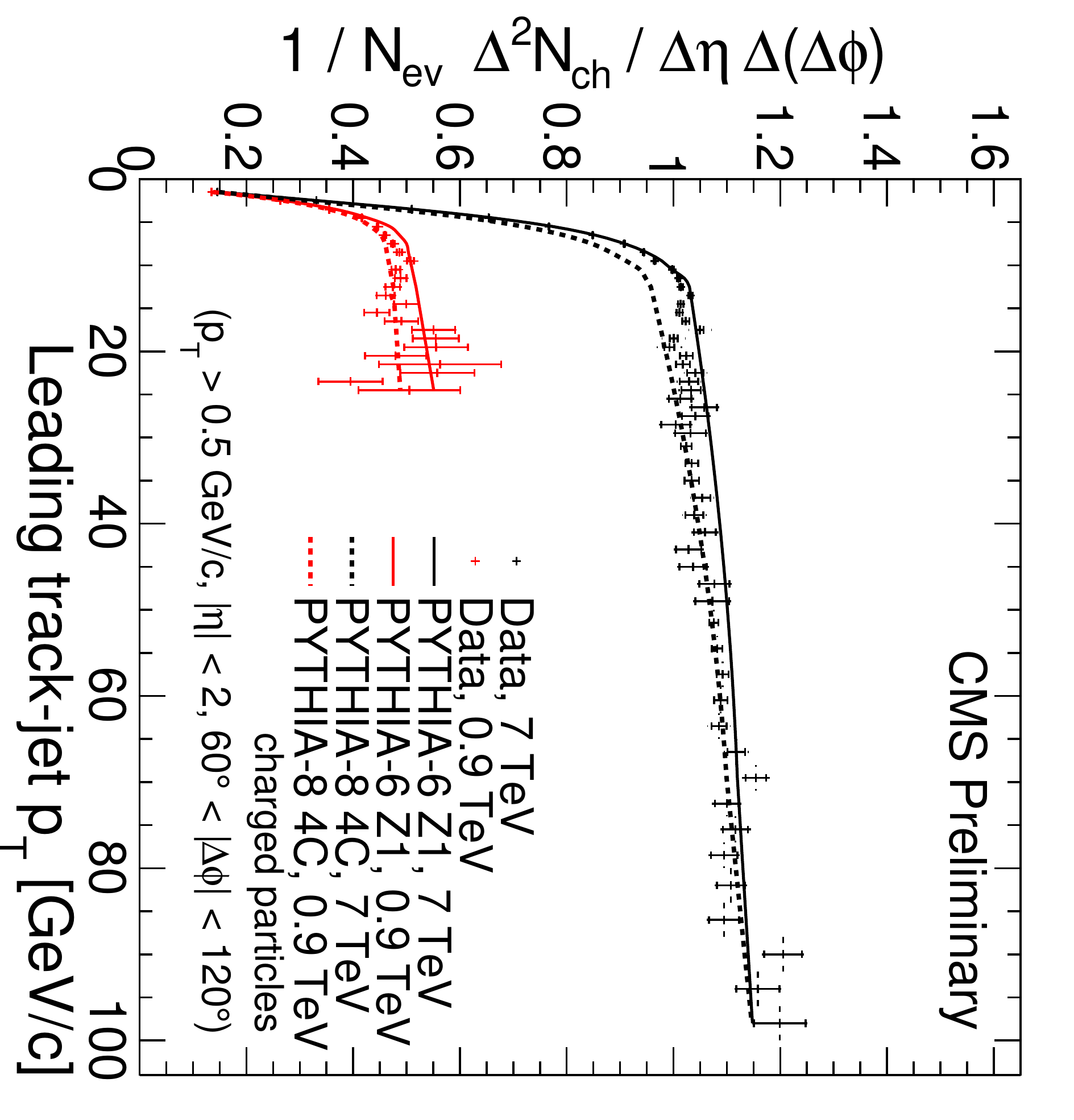}
\includegraphics[angle=90, width=0.4 \textwidth]{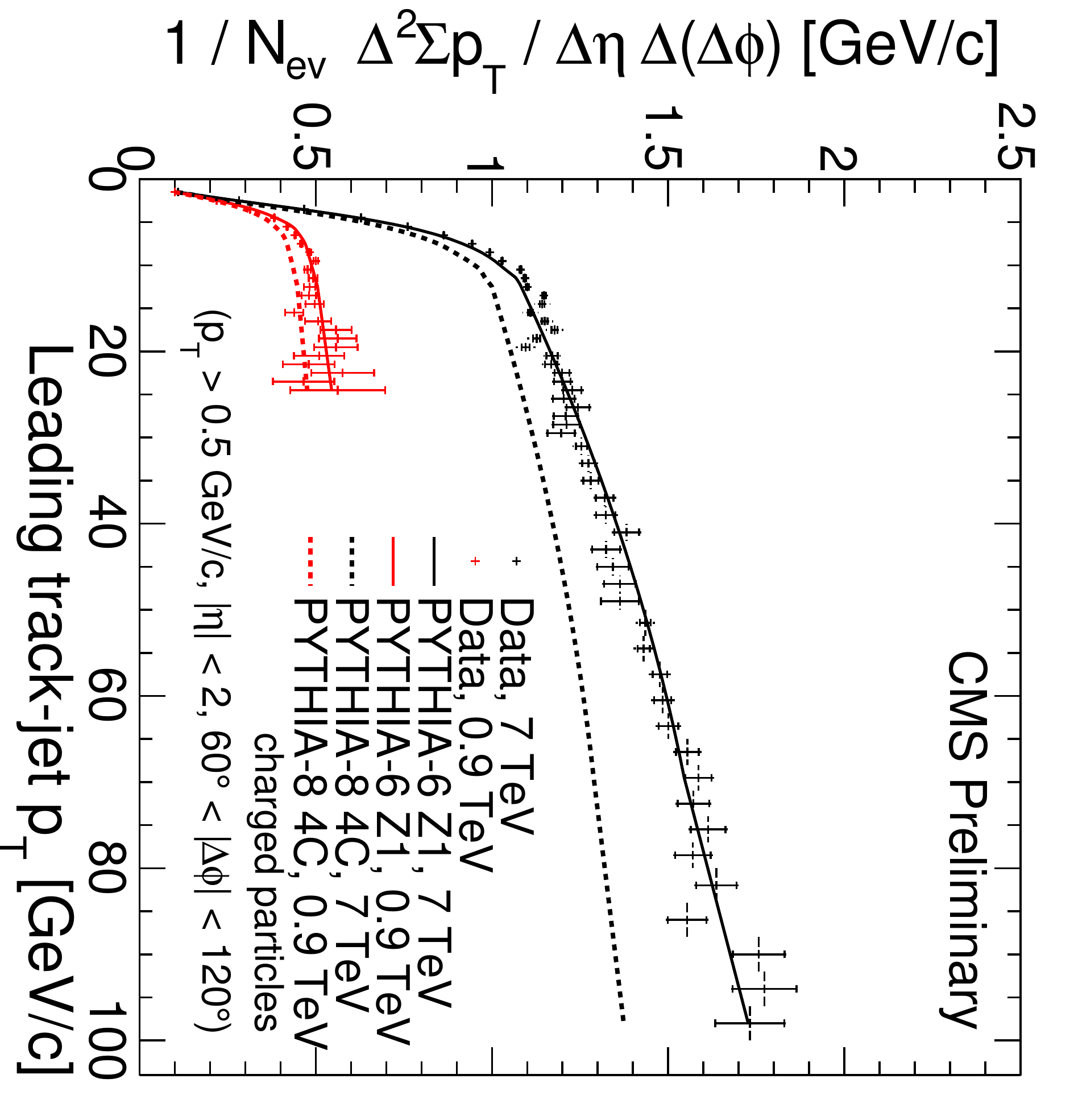} 
\includegraphics[angle=0, width=0.40 \textwidth]{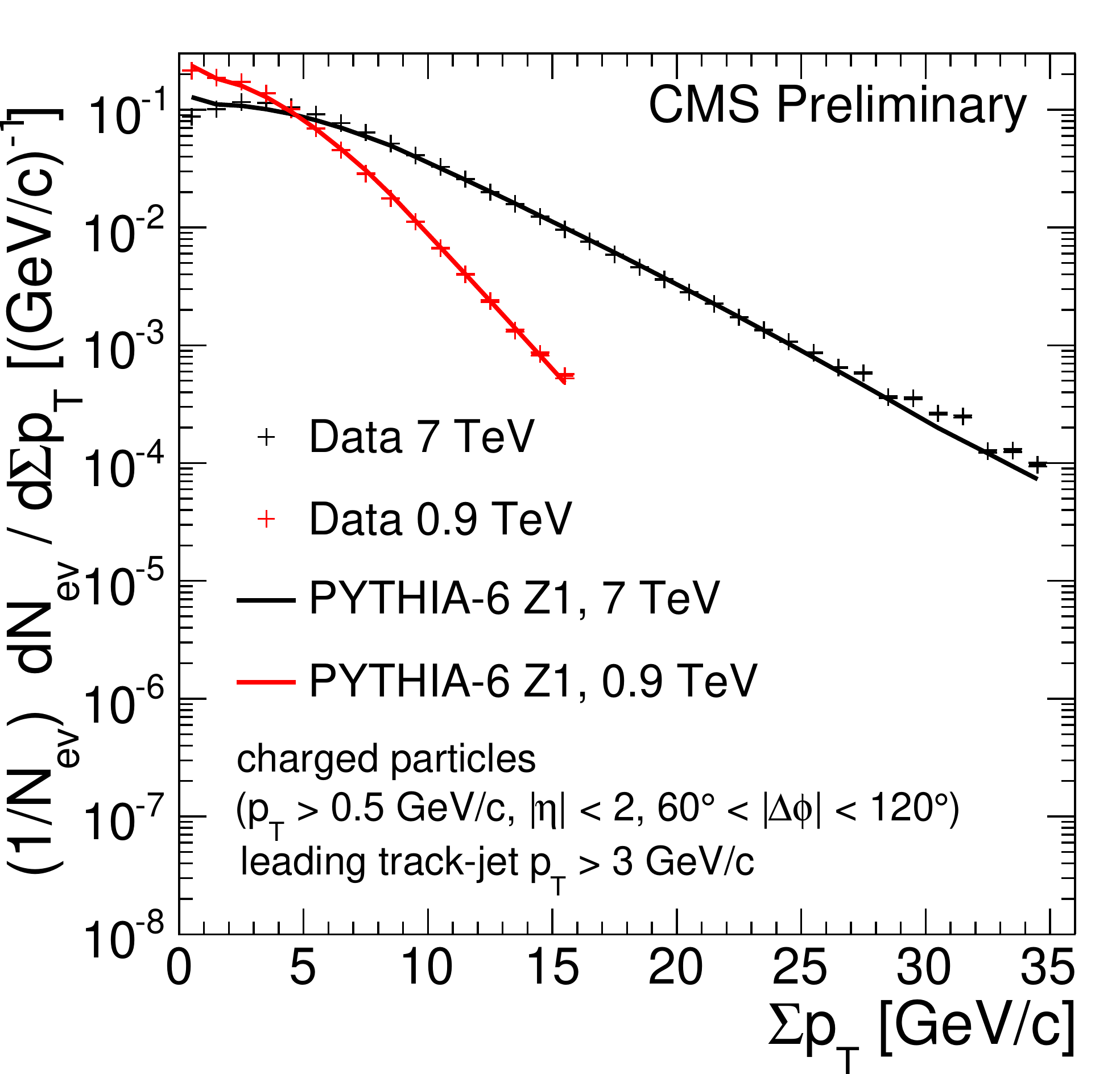} 
\includegraphics[angle=0, width=0.41 \textwidth]{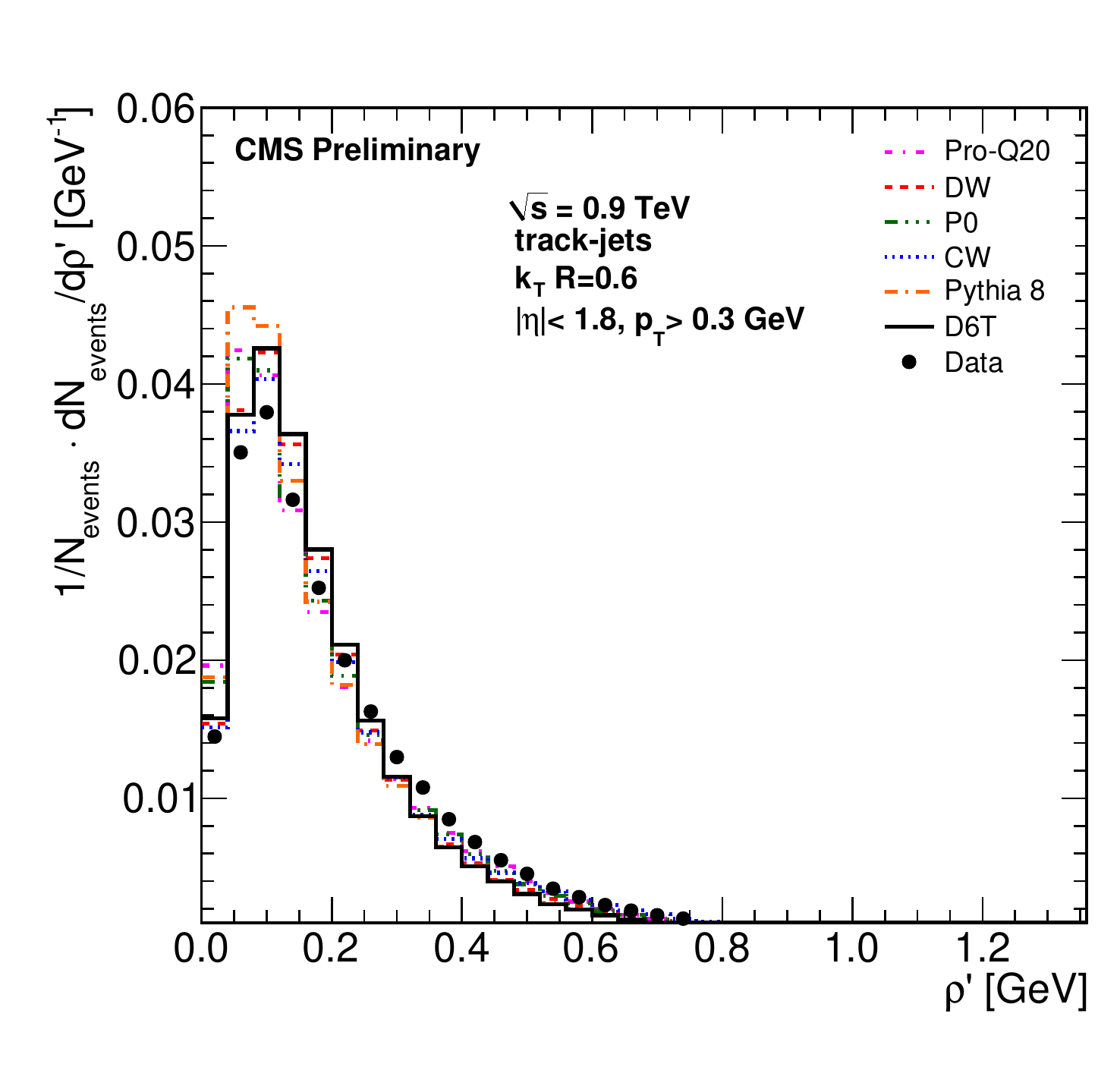} 
\end{center}
\caption{ 
(Upper plots) average multiplicity and average scalar $\sum p_T$ in the transverse region as a function of the leading track-jet $p_T$, for
data at $\sqrt{s} = 0.9$ TeV and $\sqrt{s} =7$ TeV.
(Bottom left plot)~normalized scalar $\sum p_T$ distribution in the transverse region for
data at $\sqrt{s} = 0.9$ TeV and $\sqrt{s} =7$ TeV; the leading track-jet is required to have $p_T > 3$ GeV/c.
Predictions from  PYTHIA-6 tune Z1 and PYTHIA-8.135 Tune 4C are compared to the corrected data. 
The inner error bars indicate the statistical uncertainties affecting the measurements, the outer error bars thus represent the statistical uncertainties on the measurements and the systematic uncertainty affecting the MC predictions added in quadrature.
(Bottom right plot)~normalized median of $p_T$ over jet area for track-jets reconstructed from
collision data at $\sqrt{s} = 0.9$ TeV~(black circles).
Predictions from several PYTHIA-6 tunes and PYTHIA-8 Tune 1 are compared to data. 
\label{fig:three-prime}}
\end{figure}
%%%%%%%%%%%%%%%%%%%%%%%%%%%%%%%%%%%%%%%%%%%%%%%

The centre-of-mass energy dependence of the hadronic activity in the transverse 
region is presented on the two top Figures~\ref{fig:three-prime} as a function of the $p_T$ of the leading track-jet.
The data points represent the average multiplicity and average scalar track-$p_T$ sum dependence, for $\sqrt{s} = 0.9$ TeV and $\sqrt{s} = 7$ TeV using tracks with a pseudorapidity $|\eta| <  2.0$ and $p_T > 0.5$ GeV/c.
A significant growth of the average multiplicity and of the average scalar $p_T$ 
sum of charged particles transverse to that of the leading track-jet is observed with increasing 
scale provided by the leading track-jet $p_T$, followed by saturation at large 
values of the scale (more evident for multiplicity profile than average scalar $p_T$ sum). 
A significant growth of the activity in the transverse region is also observed,
for the same value of the leading track-jet $p_T$, from $\sqrt{s} = 0.9$ TeV
to $\sqrt{s} = 7$ TeV.
These observations are consistent with the ones obtained at Tevatron \cite {Affolder:2001xt}.
The evolution with the hard scale of the ratio of the UE activity at 7 TeV and 0.9 TeV is remarkably well described by the Z1 MC. The trend is also reproduced by PYTHIA-8 Tune 4C. 
The Z2 predictions at $\sqrt{s}= 0.9$ TeV (not shown here) agree with Z1 in shape, but the normalization is 5-10\% higher for both the observables; this trend is opposite with respect to the one observed at 7 TeV and indicates that a less pronounced $\sqrt{s} $ dependence of the transverse momentum cut-off should be adopted for tunes using the CTEQ6LL PDF set than for the tunes optimized for the CTEQ5L set.

The strong growth of UE activity with	 charged particles is also striking in the comparison of the normalized distributions of charged particle multiplicity (not shown here) and of scalar $p_T$ sum which is presented in bottom-left plot of Figure~\ref{fig:three-prime} for events at	$\sqrt{s} = 0.9$ TeV and $\sqrt{s} = 7$ TeV with leading track-jet $p_T > 3$ GeV/c. The particle $p_T$ spectrum (not shown) extends up to $p_T > 10$ GeV/c, indicating the presence of 
a hard component in particle production in the transverse region.
The distributions for track-jet $p_T > 3$ GeV/c, which extend up to quite large values of the selected observables in the transverse region 
are quite well described by the various MC models, over several orders of 
magnitude. This observation gives support to the implementation of MPI in PYTHIA.
%~\cite{AAA-QCD-10-010}.

%%% QCD-10-005
The novel technique to quote the UE activity~\cite{Cacciari:exp} relies on the introduction of ``ghosts'', virtual deposits of very low energy filling the overall phase space which are taken into account by the jet clustering algorithm. 
The estimator of the overall soft background activity in an event can be derived as the median of the ratio between the transverse momentum and the area of the jets. One of the advantages of using the median compared to the mean is that it turns out to be less sensitive to the influence of outliers, i.e.\ in particular the leading jets in an event.
In order to cope with the low occupancy observed at $\sqrt{s} = $ 0.9 TeV, CMS redefines such observable restricting the median only to those jets which have physical deposits on top of ghosts:
\begin{equation}
  \rho'=
  {\mathrm{median}}\left[\left\{{\frac{p_{\textrm{T}j}}{A_j}}\right\}\right] \cdot C
\end{equation}
where $C$ is the occupancy of the event, which is the summed area $\sum_{j}{A_{j}}$ covered these jets divided by the considered detector region $A_{\rm tot}$. 
In the CMS analysis at $\sqrt{s} = 0.9$ TeV, jets are reconstructed with the anti-$K_T$ algorithm using tracks with $|\eta| <  2.0$ and $p_T > 0.3$ GeV/c.  In the right bottom plot of Figure~\ref{fig:three-prime} the $\rho'$ observable is presented for minimum bias events. The general pattern of deviations from data with respect to the considered PYTHIA tunes looks rather similar to the one observed with the traditional UE measurement. 

%%%%%%%%%%%%%%%%%%%%%%%%%%%%%
% UE IN FWD REGION
%%%%%%%%%%%%%%%%%%%%%%%%%%%%%
\section{Study of the Activity in the Forward Region}

CMS reports a measurement of the energy flow in the forward region ($ 3.15<|\eta|< 4.9$, where $\eta$ denotes the pseudorapidity) \cite{FWD-10-002} for minimum bias and dijet events in pp interactions with centre-of-mass energies $\sqrt{s}$ of 0.9 TeV, 2.36 TeV and 7 TeV. This measurement is connected to the ones reported in the previous sections as the basic philosophy is the same: it concentrates on the complementary activity of a pp interaction for different energy scales of the reconstructed leading objects.

\begin{figure}[htbp]
\begin{center} 
\includegraphics[angle=90, width=0.435 \textwidth]{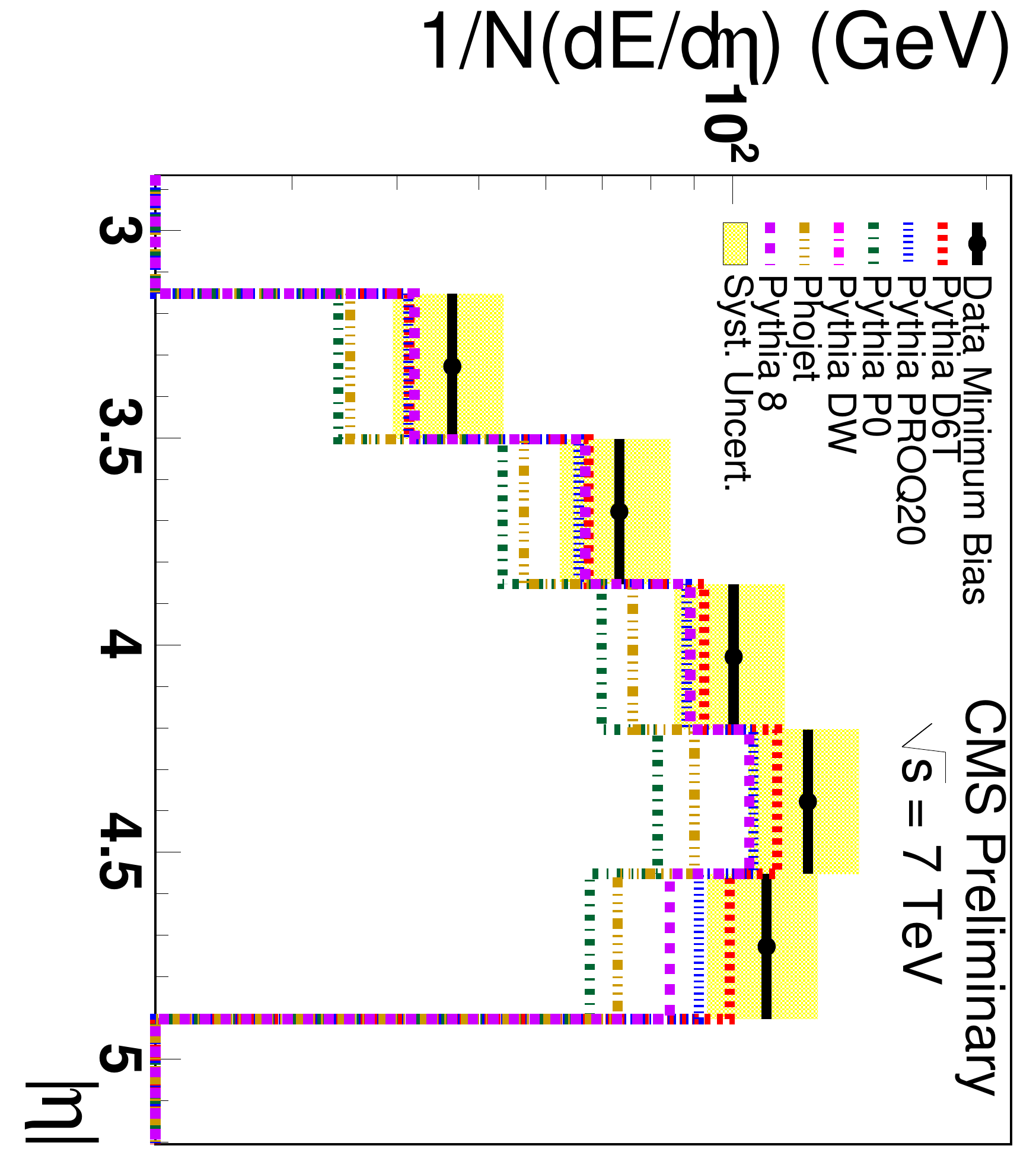}
% \hspace{2cm}
\includegraphics[angle=90, width=0.4 \textwidth]{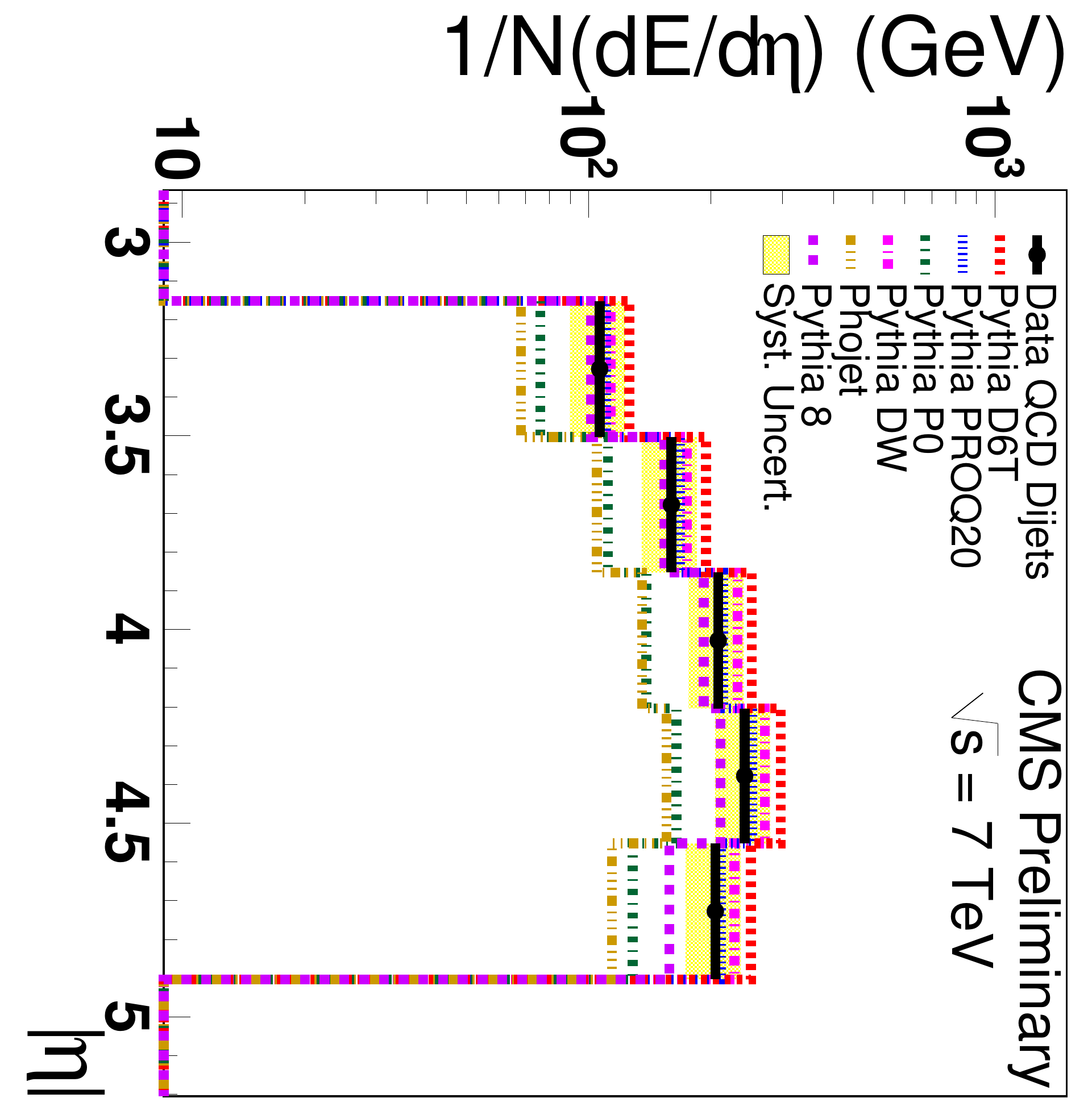}
\caption{Energy flow in the minimum bias (left) and di-jet (right) samples as a function of $\eta_{obs}$ at $\sqrt{s}=7$ TeV. Uncorrected data are shown as points, the histograms correspond to the MC predictions. Error bars corresponds to statistical errors. The shaded bands in these plots represent the systematic uncertainties of the measurements, which are largely correlated point-to-point.}
\label{resultsMBMCE}
\end{center}
\end{figure}

The energy flow in the region of the Hadron Forward detector is measured in two different event classes: in minimum bias events and in events with a  hard scale provided by a dijet system at central pseudorapidities ($|\eta| < 2.5$) and with transverse energy $E_{T,jet}>8$~GeV for $\sqrt{s} = $ 0.9~TeV and 2.36~TeV; the dijet threshold is increased to $20$~GeV for $\sqrt{s} =$ 7~TeV.
The results are qualitatively similar at all the investigated centre of mass energies.
Fig.~\ref{resultsMBMCE} shows the results of the forward energy flow at $\sqrt{s} =$ 7~TeV for the two event classes compared to predictions from Monte Carlo event generators.
The measured forward energy flow is found to be significantly different between the two event classes, with a sensitive increase and a more central activity seen in dijet events. 

\section{Multiple Parton Interactions in High-$p_T$ Phenomenology \label{feas}}

Quantifying the MPI cross sections basically deals with the measurement of $\sigma_{eff}$, the scale factor which characterizes the inclusive rate of the interactions~\cite{daniele1, daniele2}. From a phenomenological point of view $\sigma_{eff}$ is a non perturbative quantity related to the transverse size of the hadrons and has the dimensions of a cross section. The measurements performed by the AFS, CDF and D0 collaborations~\cite{Akesson:1986iv,Alitti:1991rd,Abe:1993rv,Abe:1997xk,D0-DS} favor smaller values of $\sigma_{eff}$ with respect to the naive expectations. The consequent increased rates of multiple parton interactions can be interpreted as an effect of the hadron structure in transverse plane~\cite{daniele3}. 
Extending such measurements at the LHC and studying the possible scale dependency of $\sigma_{eff}$ is definitely of great interest and may have a deep impact on the data driven estimations of the MPI backgrounds to new physics.

The production of four high-$p_T$ jets is the most prominent process to search for multiple high $p_T$ scatterings: two independent scatters in the same $pp$ or $p\bar{p}$ collision (\emph{double-parton scattering, DPS}) each producing two jets. Such a signature has been searched for by the AFS experiment at the CERN ISR, by the UA2 experiment at the CERN S$\bar{p}p$S and by the CDF and D0 experiment at the Fermilab Tevatron. 

\begin{figure}[htpb]
\begin{center}
\includegraphics[width=0.28\columnwidth]{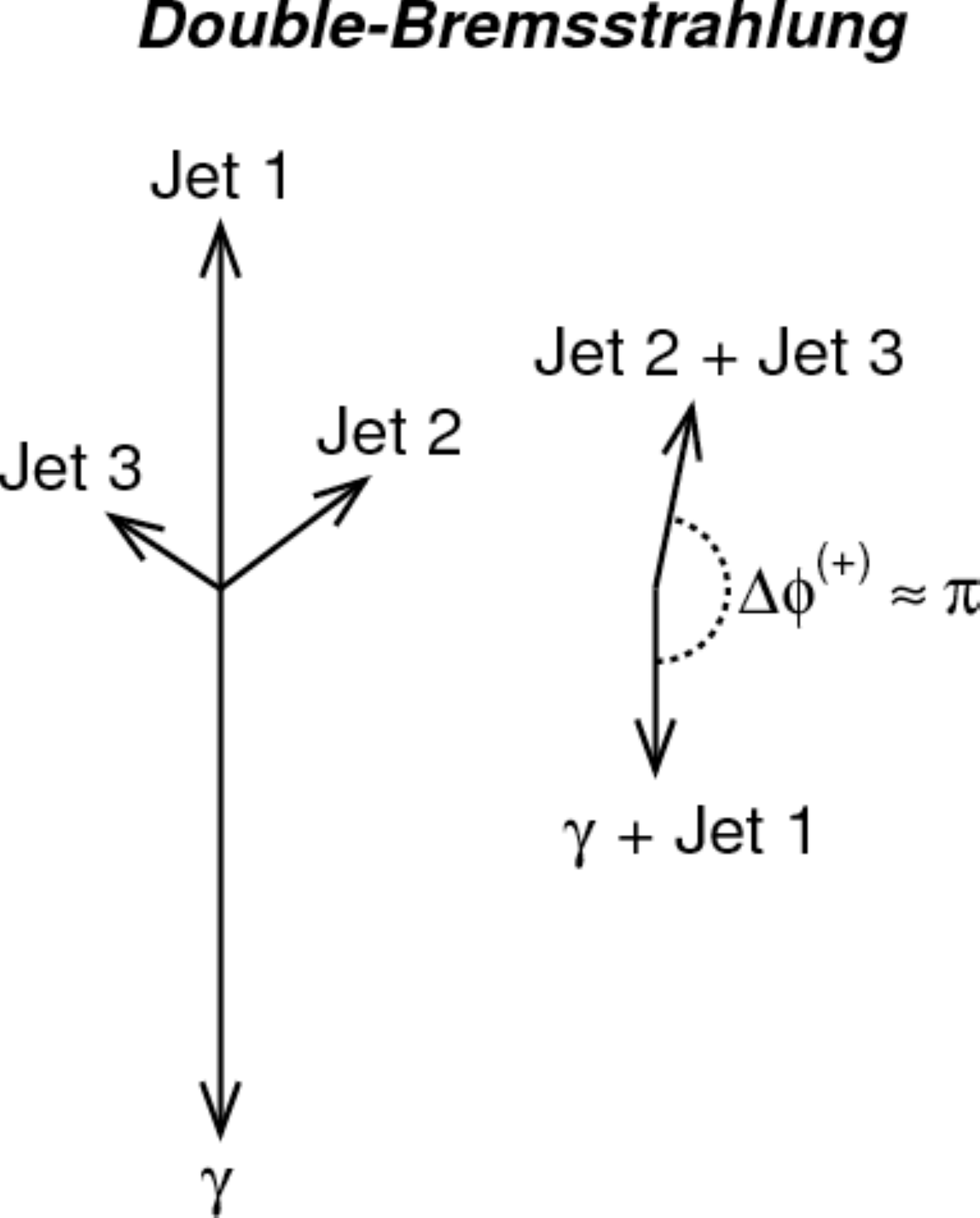} \ \ \ \ \ %
\includegraphics[width=0.28\columnwidth]{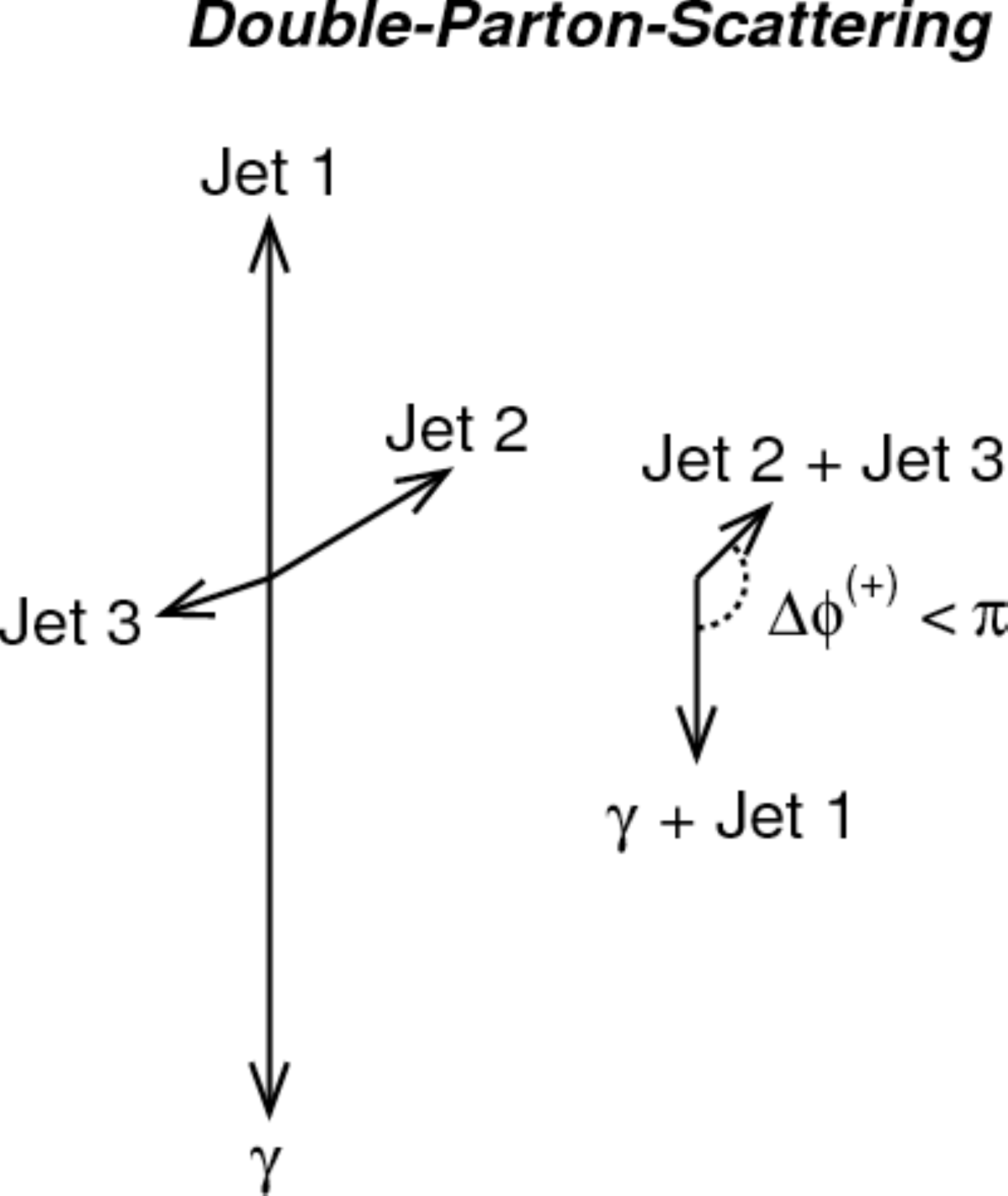}
\caption{Definition of azimuthal angle between pairs, together with typical configurations of double-bremsstrahlung (left) and double-parton scattering events (right).}
\label{fig:deltaS-sketch}
%\end{wrapfigure}
\end{center}
\end{figure}

However, searches for double-parton scattering in four-jet events at hadron colliders may face significant backgrounds from other sources of jet production, in particular from QCD brems\-strahlung (Fig.~\ref{fig:deltaS-sketch}-left). Typical thresholds employed in jet triggers bias the event sample towards hard scatterings. However, a high-$p_T$ jet parton is more likely to radiate additional partons, thus producing further jets. Thus, the relative fraction of jets from final-state showers above a given threshold is enlarged in jet trigger streams which is an unwanted bias. On the other hand, looking for four jets in a minimum-bias stream will yield little statistics. 

\begin{figure}[htpb]
  \includegraphics[width=0.5\columnwidth]{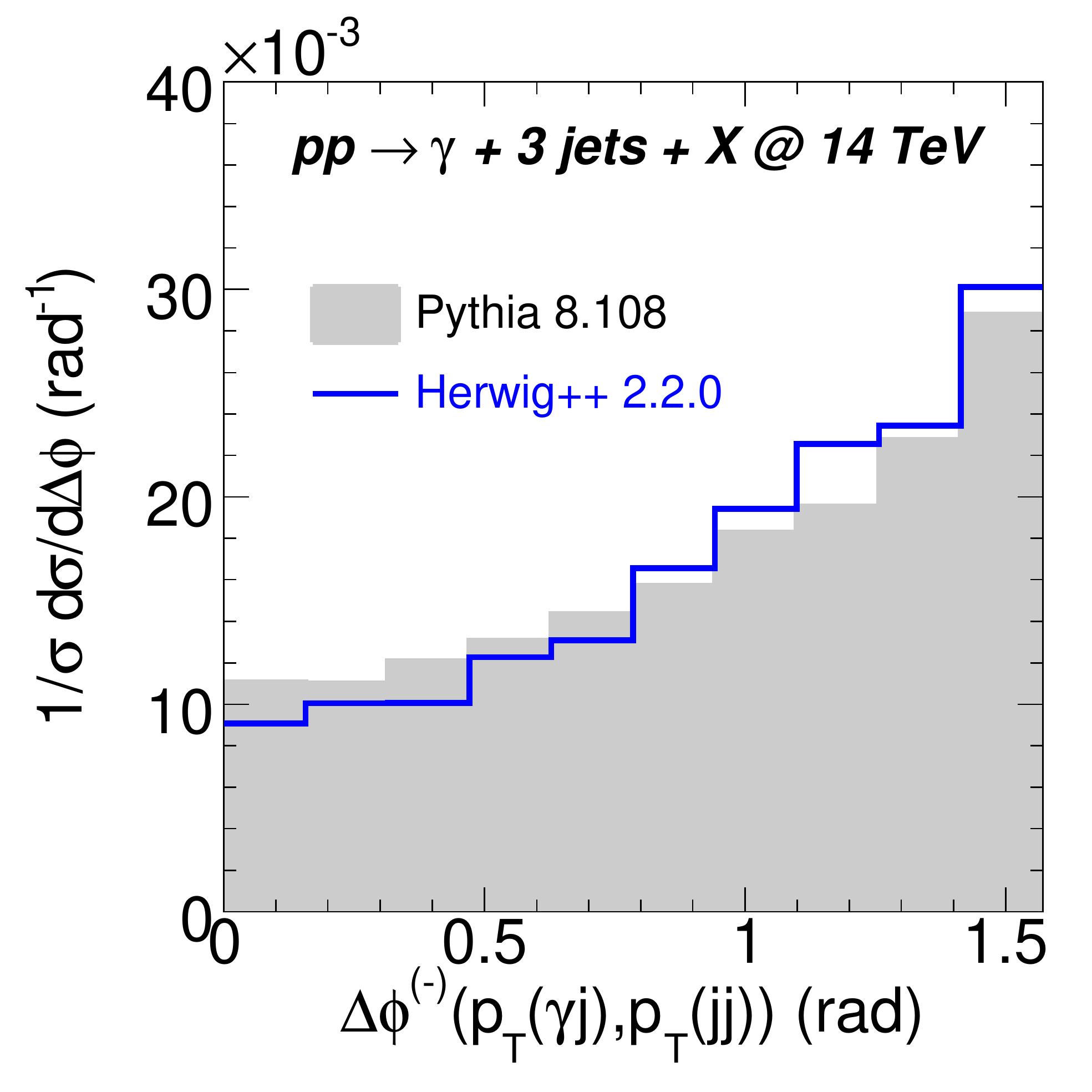} %
  \includegraphics[width=0.5\columnwidth]{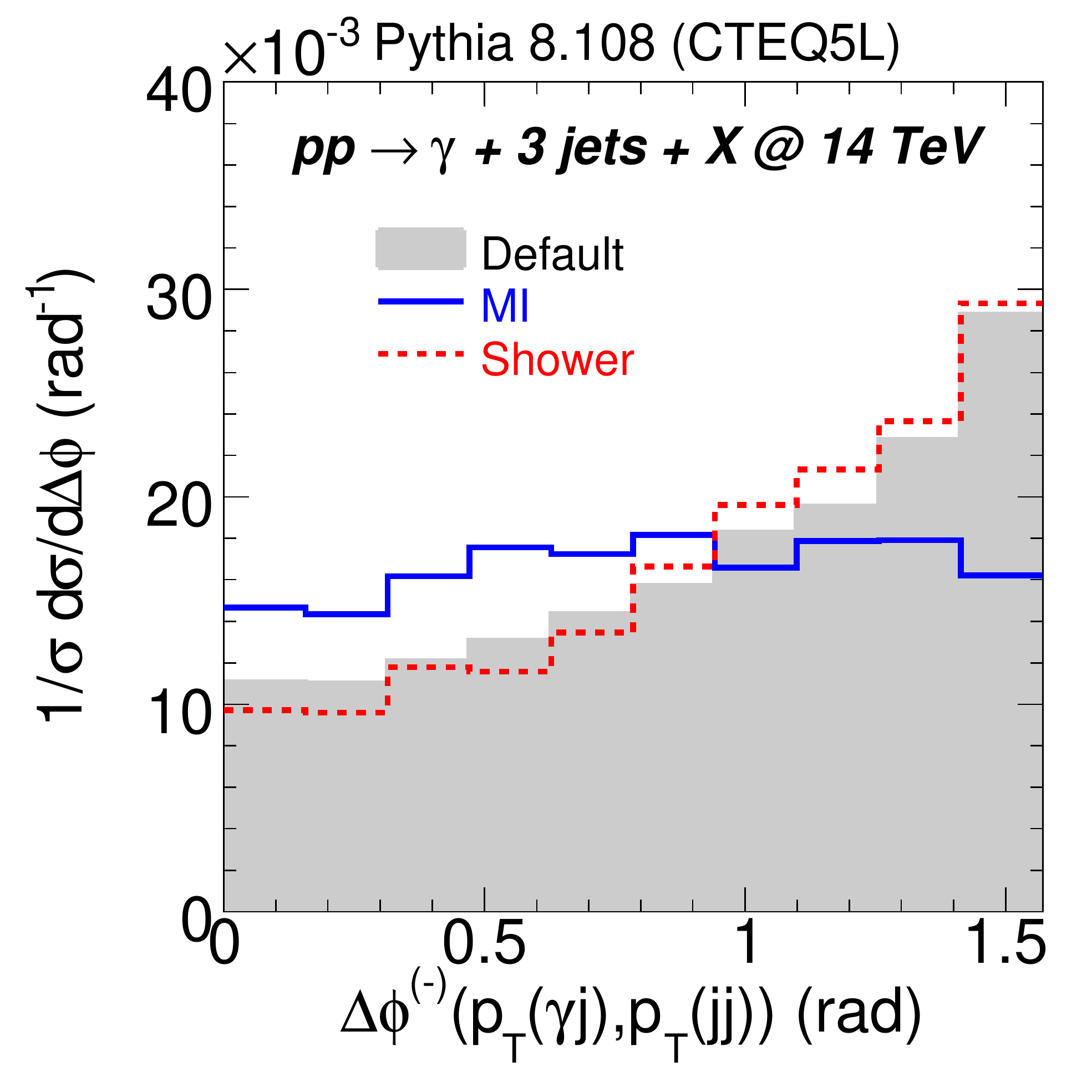} 
  \includegraphics[width=0.5\columnwidth]{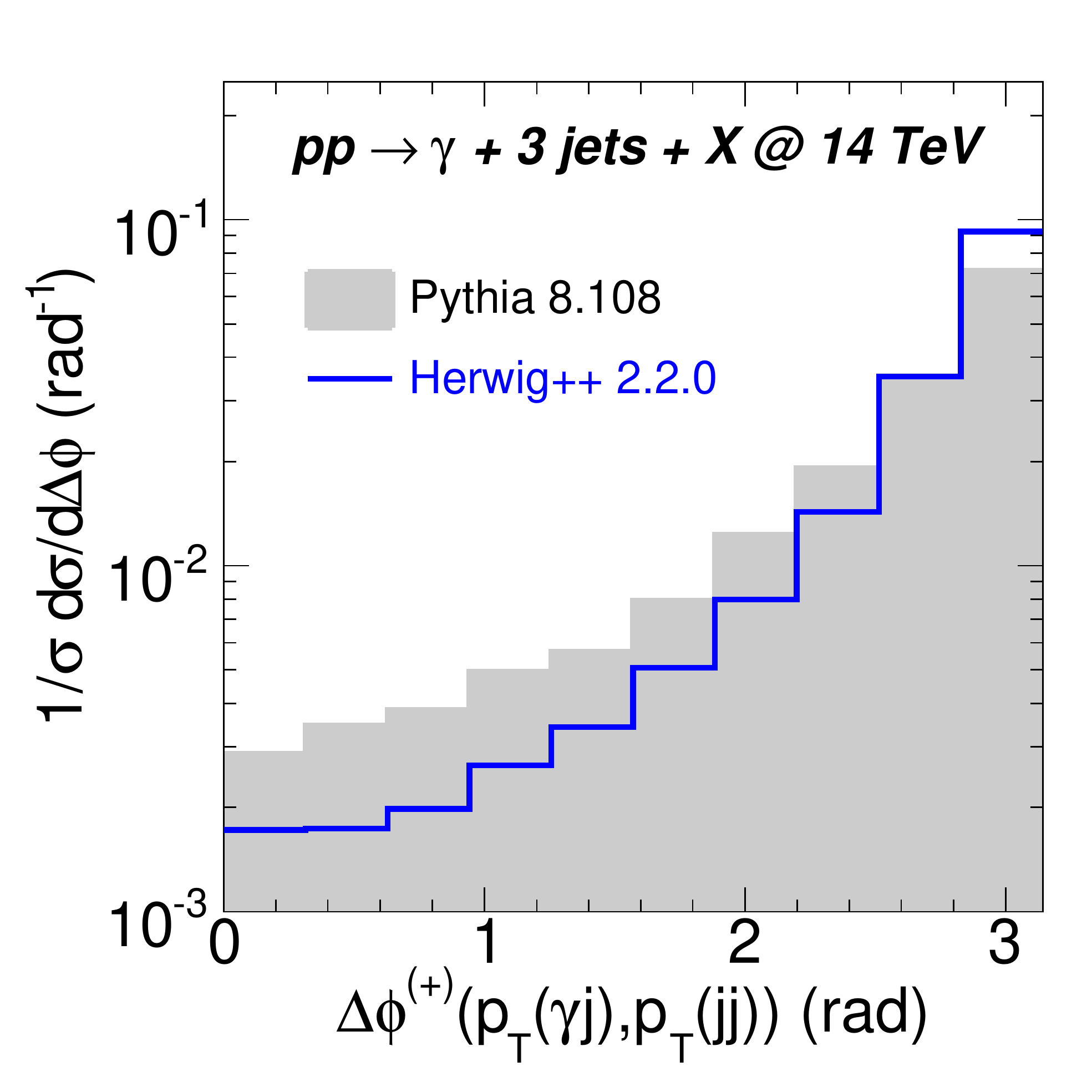}%
  \includegraphics[width=0.5\columnwidth]{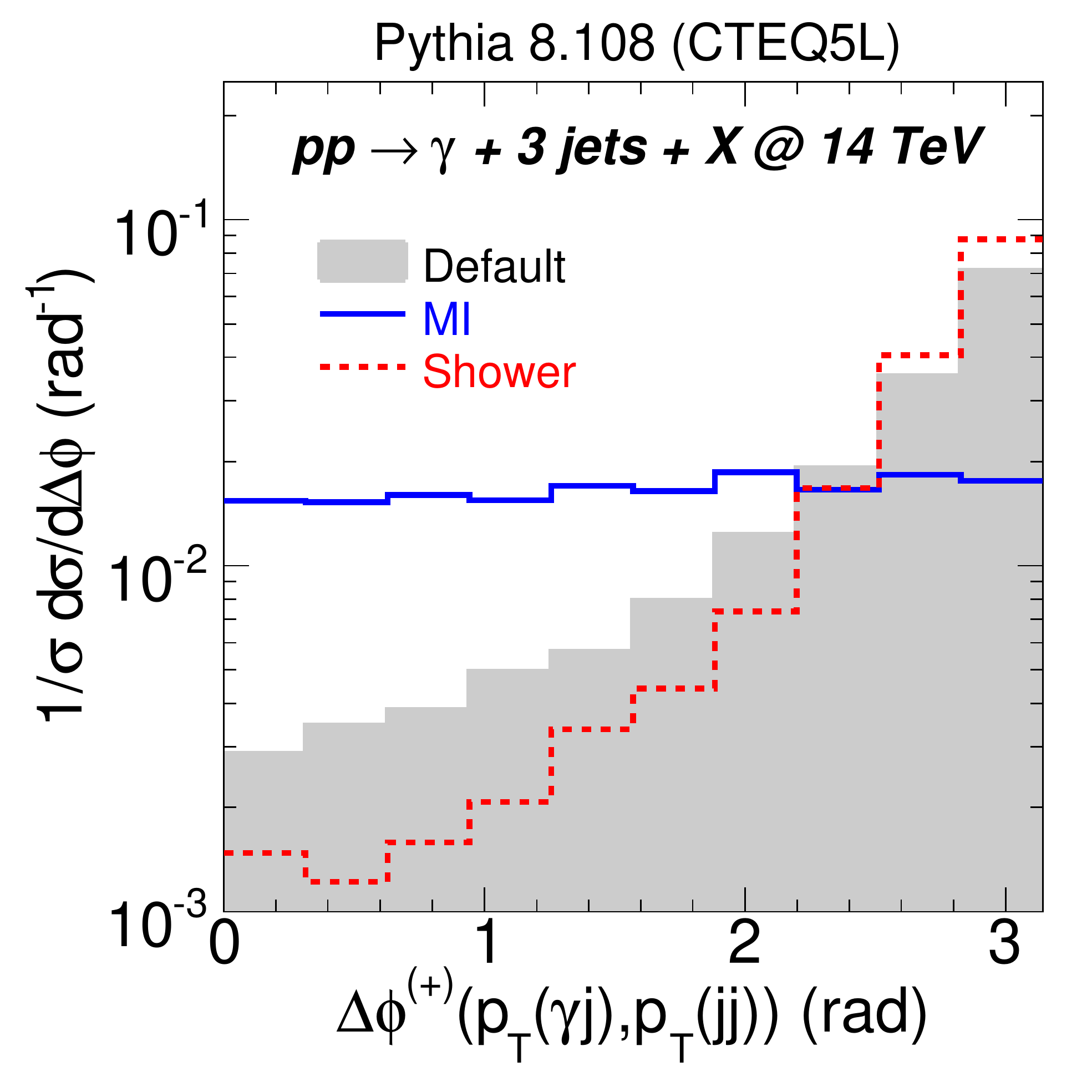}
  \caption%
  {Differential cross section shape as a function of $\Delta \phi^{(-)}$ (upper plots) and $\Delta \phi^{(+)}$ (bottom plots) variables. Predictions from PYTHIA 8.108 (\emph{Default} scenario) and HERWIG 2.2.0 (left panel) and from three different {\sc Pythia} settings (right panel) shown.}
  \label{fig:dphi}
\end{figure}
Therefore the strategy to directly measure the MPI rate in high-$p_T$ regime at hadron colliders also includes the study of multi-jets or jets+photon final states. Indeed the CDF and D0 collaborations studied final states with one photon and three jets looking for pairwise balanced photon-jet and dijet combinations. The data sample was selected with the experiment's inclusive photon trigger, thereby avoiding a bias on the jet energy. The better energy resolution of photons compared to jets purifies the identification of $E_T$ balanced pairs. Tevatron found an excess in pairs that are uncorrelated in azimuth with respect to the predictions from models without additional hard parton scatters per proton-proton scatter. CDF interpreted this result as an observation of double-parton-scatters.

Analyses trying to identify two hard scatters in multi-jet events typically rely on methodologies which overcome combinatorics. There are three possible ways to group four objects into two pairs: combinations are commonly selected pairing objects which are balanced in azimuth and energy. The flavor or other specific features of the jets may be used to decrease the combinatorics and to make looser the constraints on the balancing. One example of such a final state is constituted by events with two $b$ jets and two additional light jets. 

In order to discriminate double-parton scatters against double-brems\-strah\-lung events, CMS studies 
the observables $\Delta \phi^{(-)}$, employed by AFS, and $\Delta \phi^{(+)}$, employed by CDF, probing the azimuthal angle between pairs (Fig.~\ref{fig:deltaS-sketch}).
Expectations for the above described variables are therefore $\Delta \phi^{(-)} \approx \pi / 2$ and $\Delta \phi^{(+)} \approx \pi$ if additional jets come from double-bremsstrahlung. Otherwise, i.~e.~if additional jets come from multiple interactions, both variables should be distributed uniformly.

Differential cross section shape predictions for the $\Delta \phi$ observables in pp interactions at $14$ TeV are shown in Fig.~\ref{fig:dphi}. HERWIG 2.2.0 \cite{Corcella:2002jc,Bahr:2008pv} and PYTHIA-8.108 with default settings which include multiple interactions and showering predict similar shapes (Fig.~\ref{fig:dphi}-left). The discrimination power of the selected observables to Multiple Parton Interaction patterns is clearly shown in Fig.~\ref{fig:dphi}-right, where events with MPI switched off (\emph{Shower} scenario) are compared to events with parton shower switched off (\emph{MI} scenario). The differences are particularly pronounced when selecting the $\Delta \phi^{(+)}$ observable. 

%With multiple interactions switched off, $\Delta \phi$ is indeed most likely to be $\Delta \phi^{(-)} \approx \pi / 2$ ($\Delta \phi^{(+)} \approx \pi$). However, the correlation is weak with a factor of $3$ between first bin and last bin, i.~e.~between events with both pairs aligned in azimuth and events with pairs orthogonal in azimuth. In fact, the difference between PYTHIA's \emph{Default} and \emph{Shower} scenarios is not significant within the available statistics (Fig.~\ref{fig:dphi}-right).  Yet, both pairs are more or less uncorrelated if additional jets come from multiple interactions (\emph{MI} scenario, Fig.~\ref{fig:dphi}-right). 

\section{Conclusions}

A strong growth of the UE activity is observed with increasing leading track-jet $p_T$ for both  $\sqrt{s} = 7$ TeV and $\sqrt{s} = 0.9$ TeV.
At 7 TeV this fast rise is followed above $\sim$8 GeV/c by a saturation region with nearly constant multiplicity and small $p_T$ increase. The same pattern is observed at 0.9 TeV, with the saturation region starting at $\sim$4 GeV/c.
A strong growth of the activity is also observed with increasing centre-of-mass energy.
The large increase of activity in the transverse region is also observed in the $\sum p_T$ distribution, indicating the presence of a hard component in the transverse region. Very good post-LHC MC tunes are available for the description of the UE in the central region.

A measurement of the underlying event using the jet-area/median approach is also reported, demonstrating its sensitivity to different underlying event scenarios. 

Complementary underlying event measurements in the forward region are also presented. The energy flow in the forward direction is measured for minimum bias and central di-jets events. A more global UE description including both the central and the forward regions is certainly one of the next MC tuning challenges, with deep impact on the understanding of the MPI dynamics.

%The predictions of several tunes have been compared to corrected DATA. These models differ in the PDF description, in the implementation of radiation, fragmentation, and the $\sqrt{s}$ dependence of the amount multiple parton interactions. A good descriptions of most distributions at $\sqrt{s} = 7$ TeV and of the $\sqrt{s}$ dependence from 0.9 to 7 TeV is provided by the Z1 tune.

The Multiple Parton Interactions measurement strategy in the high-$p_T$ regime is also briefly discussed focusing on the $3jet + \gamma$ topology. The very good performances of the LHC machine should allow to have soon the integrated luminosity conditions adequate to perform these measurements over a wide range of energy scales, with deep impact on the data driven estimation of the MPI backgrounds to searches.

\end{document}